\documentclass[useAMS,usenatbib]{mn2e}
\usepackage{epsfig}
\newcommand{\ISO}{{\em ISO}}
\newcommand{\DIRBE}{{\em DIRBE}}
\newcommand{\IRAS}{{\em IRAS}}

\begin{document}


\title[ELAIS VIII: 90\,$\mu$m source counts]{The European Large
 Area \ISO\ Survey  VIII: \\90\,$\mu$m  final analysis and source counts}

\author[Philippe H\'eraudeau et al.]
{\parbox{159mm}{\begin{flushleft}
{\LARGE Ph.~H\'eraudeau$^{1,2}$\thanks{E-mail:
P.Heraudeau@astro.rug.nl}, }
{\LARGE S.~Oliver$^{3}$, } 
{\LARGE C.~del Burgo$^{4,2}$, } 
{\LARGE C.~Kiss$^{5,2}$, } 
{\LARGE M.~Stickel$^{2}$, } 
{\LARGE T.~Mueller$^{6}$, } 
{\LARGE M.~Rowan-Robinson$^{7}$, }
{\LARGE A.~Efstathiou$^{8}$,  }
{\LARGE C.~Surace$^{9,7,2}$,}
{\LARGE L.V.~T\'oth$^{5,2}$,}
{\LARGE S.~Serjeant$^{10}$,}
{\LARGE D. M. Alexander$^{11}$,}
{\LARGE A.~Franceschini$^{12}$,}
{\LARGE D.~Lemke$^{2}$,}
{\LARGE T.~Morel$^{13}$,}
{\LARGE I.~P\'erez-Fournon$^{14}$,}
{\LARGE J-L.~Puget$^{15}$,}
{\LARGE D.~Rigopoulou$^{16}$,}
{\LARGE B.~Rocca-Volmerange$^{17}$,}
{\LARGE A.~Verma$^{6}$}
\end{flushleft}
}\vspace*{0.200cm}\\
\parbox{159mm}{
$^{1}$ Kapteyn Astronomical Institute, Postbus 800, 9700 AV Groningen, The Netherlands\\
$^{2}$ Max-Planck-Institut f\"{u}r Astronomie, K\"{o}nigstuhl 17, D-69117, Heidelberg, Germany\\
$^{3}$ Astronomy Centre, University of Sussex, Falmer, Brighton BN1 9QJ\\
$^{4}$ European Space \& Technology Centre (ESTEC), Keplerlaan 1,  Postbus 299, 2200
AG Noordwijk, The Netherlands \\
$^{5}$ Konkoly Observatory, P.O. Box 67., H-1525 Budapest, Hungary\\ 
$^{6}$ Max-Planck-Institut f\"{u}r extraterrestrische Physik, Giessenbachstra{\ss}e, 85748 Garching, Germany \\
$^{7}$ Astrophysics Group, Blackett Laboratory, Imperial College of
Science Technology \& Medicine, Prince Consort Rd., London SW7 2BZ\\
$^{8}$ Department of Computer Science and Engineering, Cyprus College, 6 Diogenous Street, P.O. Box 22006, 1516 Nicosia, Cyprus \\
$^{9}$ Laboratoire d'Astrophysique de Marseille, Traverse  du Siphon-Les
trois Lucs, BP8, F13376 Marseille Cedex 12 \\
$^{10}$ Unit for Space Sciences \& Astrophysics, School of Physical Sciences,
University of Kent at Canterbury, Canterbury, Kent CT2 7NZ\\
$^{11}$ Institute of Astronomy, Madingley Road, Cambridge, CB3 0HA\\
$^{12}$ Dipartimento di Astronomia, Universita' di Padova, Vicolo Osservatorio 5, I-35122 Padova, Italy\\
$^{13}$ Osservatorio Astronomico di Palermo, Piazza del Parlamento 1, I-90134 Palermo, Italy\\
$^{14}$ Instituto de Astrofisica de Canarias, C/V\'\i a L\'actea s/n, 38200 La Laguna, Tenerife, Spain \\
$^{15}$ Institut d'Astrophysique Spatiale,  B\^{a}timent 121, Universit\'{e} Paris XI, 91405 Orsay cedex, France\\
$^{16}$ Department of Physics, Denys Wilkinson Bldg., University of Oxford, Keble Road, Oxford, OX1 3RH \\
$^{17}$ Institut d'Astrophysique de Paris, 98bis Boulevard Arago, F 75014 Paris, France
}}

\date{Submitted ; accepted }

\pagerange{\pageref{firstpage}--\pageref{lastpage}} \pubyear{2002}

\maketitle

\label{firstpage}

\begin{abstract}
We present a re--analysis of the European Large Area \ISO\ Survey (ELAIS)
90\,$\mu$m observations carried out with {\em ISOPHOT}, an instrument on board the ESA's Infrared Space 
Observatory ({\em ISO}). With more than 12\,deg$^2$, the ELAIS survey is the largest
area covered 
by \ISO\ in a single program and is about one order of magnitude deeper than the \IRAS\ 100\,$\mu$m survey.
The data analysis is presented and was mainly performed with 
the Phot Interactive Analysis software (PIA, Gabriel et al. 1997) but using the
pairwise method of Stickel et al. (2003) for signal processing from ERD 
(Edited Raw Data) to SCP (Signal per Chopper Plateau). 
The ELAIS 90\,$\mu$m catalogue contains 229 reliable sources with fluxes larger than 70
mJy and is available at http://www.blackwell-synergy.com.
Number counts are presented and show an excess 
above the no-evolution model prediction.
This confirms the strong evolution detected 
at shorter(15\,$\mu$m) and longer (170\,$\mu$m) wavelengths in other \ISO\
surveys. The ELAIS counts are in agreement with previous works at 90\,$\mu$m and
in particular with the deeper counts 
extracted from the Lockman hole observations (Rodighiero et al. 2003).
Comparison with recent evolutionary models show that the models 
 of Franceschini et al. (2001) and Guiderdoni et al. 
(1998) which includes a heavily-extinguished population of galaxies give 
the best fit to the data. 
Deeper observations are nevertheless required to better discriminate between 
the model predictions in the far-infrared and are scheduled with the Spitzer Space 
Telescope (e.g. Lonsdale et al. 2003) which already started operating and
will also be performed by ASTRO-F (e.g. Pearson et al. 2004). 

\end{abstract}

\section{Introduction}
Strong evolution has been detected in the infrared regime based on \IRAS\
number counts at 12, 25, 60 and 100\,$\mu$m (Hacking \& Houck 1987, Hacking, Condon \& Houck 1987, Hacking   \& Soifer 1991 , Oliver, Rowan-Robinson \& Saunders 1992, Bertin,  Dennefeld \& Moshir 1997) which show an excess of galaxies compared to the no-evolution scenario.
 These findings were recently confirmed with much deeper surveys carried out with 
the {\em ISOPHOT\/} instrument on-board the Infrared Space Observatory ({\em ISO})
(Kessler et al. 1996) at 90 and 170\,$\mu$m  (Kawara et
al. 1998,  Puget et al. 1999,  Efstathiou et al. 2000, Linden-V{\o}rnle et al. 2000, Juvela, Mattila
\& Lemke 2000, Matsuhara et al. 2000,  Dole et al. 2001).
\ISO\ also detected a substantial number of faint sources, consistent with strong 
evolution from 15\,$\mu$m number counts (see e.g. Elbaz et al. 1999, Gruppioni et
al. 2002). Differential counts obtained from several independent 15\,$\mu$m 
{\em ISOCAM\/} surveys show a remarkable upturn at $S_{15}<3$ mJy and an excess
of a factor 10 at the faintest flux above the no-evolution predictions.

In addition to the excess of galaxies detected by \ISO\ surveys from the
mid-infrared to the far-infrared, the observational constraints set by 
the discovery of the cosmic infrared background (CIB) (see Hauser \& Dwek 2001
for a review and references therein) together with deep submillimetre surveys 
(Hughes and Dunlop 1998, Barger et al. 1998, Eales et al. 2000, Scott et al. 2002, Webb et al. 2003) are dramatically 
increasing the development of new scenarios of galaxy formation and evolution
(Pearson \& Rowan-Robinson 1996, Guiderdoni et al. 1998, Devriendt \&
Guiderdoni 2000, Rowan-Robinson 2001, Franceschini et al. 2001, Takeuchi et al. 2001, Pearson 2001, Wang 2002, Lagache et al. 2003, Xu et al. 2003).

The ELAIS survey (for an overview see Oliver et al. 2000, Paper I) was the largest 
open time project conducted by {{\em ISO}}. 
This survey consists of more than 12\,deg$^2$
of the sky surveyed at 15 and 90\,$\mu$m, nearly 6\,deg$^2$ at 6.7\,$\mu$m
and 1\,deg$^2$ at 175\,$\mu$m 
(i.e. the FIRBACK survey, see Puget et al. 1999) in four high 
Ecliptic latitude ($|\beta |$ $>$40$^{\rm o}$) regions with low \IRAS\ 100\,$\mu$m 
sky brightness ($<$ 1.5 MJy sr$^{-1}$). 
In this work, we present a 90\,$\mu$m analysis and source counts limited to the 
four large areas, three in the northern hemisphere (N1, N2 and N3), and
one in the southern hemisphere (S1).
Preliminary results of the ELAIS survey at 90\,$\mu$m based on the Quick 
Look analysis and the brightest sources were presented in Efstathiou et 
al. 2000 (hereafter referred to as Paper III). 

The paper is organized as follows: in Section \ref{sec_obs} we describe the observations 
and the data reduction based 
on the analysis of the distribution of consecutive read-outs of the 
detector instead of using the whole ramp. After the source extraction (Section \ref{sec_extraction}), we
search in vain for solar system objects to remove them from the source list
and since some could be useful for calibration purposes.
In Section \ref{src_reliability}, we estimate the completeness of the survey, source flux and position
accuracies and the Eddington bias correction from Monte-Carlo simulations of artificial sources on the 
final maps.  
The final catalog of sources is presented in Section \ref{sec_cata}. 
We compare the {\em ISOPHOT\/} calibration for all standard stars
observed at 90\,$\mu$m with model predictions and for 
the sources detected in the survey with \IRAS\ values (Section \ref{calib_comparison}).  
Temperatures from colour ratios between 90 and 170\,$\mu$m for sources also
detected in the
FIRBACK survey are computed in Section \ref{sec_90_170}. 
After computing the structure noise (Section \ref{confusion_noise})
in the ELAIS fields we present number counts 
(Section \ref{sec_counts}) which are compared
with other works at 90\,$\mu$m and to evolutionary models before the summary
 and discussion of our results in section \ref{sec_conclusion}. 

\section{Observations and data processing}
\label{sec_obs}
\subsection{Observations}
\label{sec_obs1}
 The 90\,$\mu$m ELAIS data consist of 13 to 20 P22 staring raster maps performed with the
 $3\times3$ array detector C100 of the {\em ISOPHOT\/} instrument (Lemke et
 al. 1996, for an overview see the {\em ISOPHOT\/} Handbook by Laureijs et al. 2003)
 on board {\it ISO}. 
 The pixel size on the sky of the C100 detector
 is 43\arcsec.5 $\times$ 43\arcsec.5 and the distance between
 the pixel centers are 46\arcsec. The 
 {\em ISOPHOT\/} filter-band C$_{\rm 90}$ with a reference wavelength of 90\,$\mu$m and a width of 51\,$\mu$m
 was used. At this wavelength, the full width at half maximum (FWHM) of the beam
 profile is 50 arcsec. Each raster map covers typically 20$\times$40 arcmin$^2$.
 Table A1 in Paper I provides full details on the observations.
 The N1, N2, N3 and S1 fields cover 2.74, 2.98, 2.16 and 4.15\,deg$^2$ on the
 sky respectively i.e. 12.03\,deg$^2$ in total.
 The exposure time was 20s but a number of sub-fields 
(representing about 17\% of the whole survey area) were re-observed 
with 12s exposures (see Fig. 14 in Paper I for the survey coverage).

\subsection{Signal processing}
 The data were first processed with the Phot Interactive Analysis (PIA)
 software (Gabriel et al. 1997)
 version 9.1 using the OLP10 calibration files  modified by the inclusion 
 of the new dark signal correction (del Burgo et al. 2003a, 2003b).
The data reduction from ERD (Edited Raw Data) to SCP (Signal per Chopper
   Plateau) was performed using the pairwise method of
   Stickel et al. (2003) which was also used by Juvela et al. (2000). 
 The signal derived from the distribution of the 
   difference between consecutive read-outs is used instead of making linear 
   fits to the whole ramps.
 After rejecting the first 10 per cent of the data stream which may be affected by
 transient, the unweighted myriad technique (Kalluri \& Arce 1998) was used
 as a robust estimator of the pairwise distribution for each raster position.
 The distribution was assumed to be Cauchy (a type of
   $\alpha$--stable distribution like Gaussians but with a heavier-tailed distribution) 
 to take into account the presence of glitches in the tail of the distribution. 

\subsection{Calibration}
\label{calibration}
 Each raster was preceded and followed by an FCS (Faint Calibration Source)
 measurement.
 However, the calibration of the on-sky measurements was made using 
 the second FCS only, performed immediately after the raster and with a power
 chosen to reproduce the intensity of the sky background of the measurement. 
 The second FCS generally shows a smaller transient behavior than the first
 one, providing a more accurate measurement.
 For each field, the relative uncertainty coming from the FCS calibration was
 computed as the mean absolute deviation of the average sky background
 of all rasters, which results to be 7 per cent.

\subsection{Flat-fielding and Mapping}

\par
Differences of up to 20\% in the overall levels of
the data streams of the detector pixels were noticed after the flux calibration. 
This behaviour is most likely resulting from pixel-to-pixel sensitivity  
differences, which moreover appeared to
be time-dependent. 
\par
To correct for this, the pixel data streams were slightly smoothed and 
filtered to remove sources. At each raster point the mean of the
filtered pixel values was computed. The sequence of ratios of the mean
and the individual pixel value at each raster point was fitted with a 
robust polynomial to give the smooth correction function for each pixel. 
If remaining time trends were still noticeable after correcting the 
individual pixels to the common
mean (by multiplication), the procedure was repeated but the filtered
data values from all detector pixels were 
simultaneously fitted with a robust low order polynomial. This removes
any time trend still present after rescaling the pixel data streams to
the common mean.

\par
This combined method is highly effective in removing
pixel-to-pixel sensitivity differences and time trends in the data
streams.
 
 The whole field map was built from the Jy/pixel values for each field using
 the drizzle mapping method under IRAF (Tody 1993) with a pixel size of 30
 arcsec 
and the default shrink factor (0.65).
 The drizzle method (Fruchter \& Hook 1997, 2002) allows to consider 
 the exact size of pixels and gaps between them (see Sect.~\ref{sec_obs1}).

\section{Source detection}
\label{sec_extraction}
 The source detection was performed using the SExtractor software
 version 2.2.2 (Bertin \& Arnouts 1996) on the final maps.
 The sky background was computed in a grid of $15\times 15$ square pixel 
 (i.e. $7.5\times7.5$ square arc-minute) and SExtractor was run with a 
 detection threshold of 1.8\,$\sigma$ and a minimum number of pixel equal to 2.
 The flux in a circular aperture (FLUX\_APER) of 6 pixels (i.e. 180 arcsec) diameter
 was used.

\subsection{Search for solar system objects in the ELAIS fields}
Although the ELAIS fields are at high ecliptic latitude (for low zodiacal
background), there could still be objects from the solar system
inside the fields. The search for these targets had two purposes:
1. Cleaning of the ELAIS source list from moving solar system
targets. 2. Finding additional targets which might be later on
used for independent flux calibration purposes (M\"uller et
al.\ 2002; M\"uller \& Lagerros 1998, 2002).
As the raster maps were observed at different periods during the \ISO\ mission,
we used the exact date and time at which they were obtained to search the databases
of the Minor Planet Center\footnote{http://cfa-www.harvard.edu/cfa/ps/mpc.html}.
For our search we included more than 150\,000 asteroids with reliable orbital
elements (numbered asteroids and unnumbered, multi-apparition objects), more than
200 comets and the planets and their satellites.
A search radius which was slightly larger than the actual ELAIS fields
was used to account for the geocentric to \ISO\-centric parallax errors (the position
calculations were done in the geocentric frame).
A geocentric to \ISO\-centric parallax of 10 arcmin covers all
asteroids beyond 0.15 AU from Earth, i.e. more than 99\% of all
known asteroids, but we allowed for parallaxes of up to 30 arcmin
to also account for possible ephemeris uncertainties and asteroid
movements during the observations. This means that the 2 deg
search radius for each ELAIS field included a very large safety
margin.

One asteroid and one comet were selected to be possibly seen in one
of the ELAIS rasters. We computed their
movements during the observation: 12 arcsec and 1.5 arcmin.
The ISO parallax correction was less than 1 arcmin in
both cases.
The two objects were finally found to be outside
the ELAIS field when we repeated the ephemeris calculation with a
more sophisticated N-body tool in the \ISO\-centric frame.
 From our analysis we concluded that there are no
known solar system sources in the ELAIS 90\,$\mu m$ data.

\section{Completeness, source flux uncertainty and position accuracy}
\label{src_reliability}
To estimate several quantities such as completeness, flux and positional
uncertainties 
we adopt a similar approach as Dole et al. (2001) for FIRBACK
based on the addition of artificial sources to the data.
 Artificial sources were added at random positions on the final map of
 each field using the 90\,$\mu$m theoretical footprint scaled by a certain factor to
 simulate sources with a known flux.

 In practice, to keep only the noise on the images, 
 objects detected by SExtractor were first removed from the images for the
 simulations (subtracting the image obtained with the SExtractor option
 CHECKIMAGE\_TYPE=OBJECTS).
 The source position can fall anywhere on a pixel and the pixelised footprint
 was computed in a square of $5 \times 5$ pixels providing a spatial extension
 of 2.5 $\times$ 2.5 arcmin$^2$ for each source and representing 96
 per cent of the total flux contained in the theoretical footprint.
 Simulations of 15 artificial sources and the extraction with SExtractor
 using the same parameters as for the survey sources were repeated 300 times for
 each field giving a total of 4500 simulated sources at each flux level equal to 50, 60, 70, 80, 90, 100, 125, 150,
 200, 300, 400, 500, 750 and 1000 mJy.
%
\subsection{Positional accuracy}
\begin{figure}
\epsfig{file=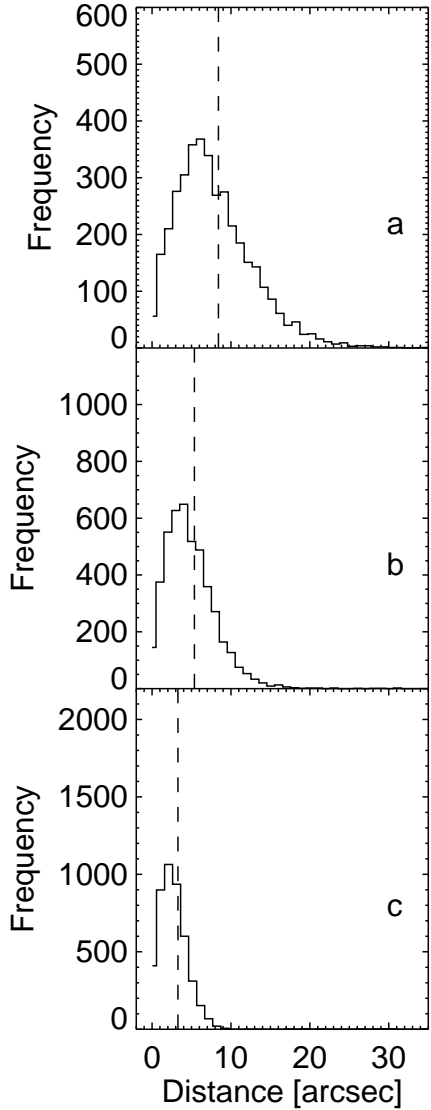,height=16cm, bb=80 80 265 540}
\caption{Distribution of the distance between recovered and simulated sources
  with a flux equal to 100, 200 and 500 mJy in plots 'a','b' and 'c'
  respectively in N2. The mean distance is indicated as a dashed line in each graph
  and is equal to 8.4, 5.4 and 3.3 arcsec for 100, 200 and 500 mJy, respectively.}
   \label{pos}
\end{figure}

 The positional accuracy can be estimated from the statistical analysis of the
 distances between recovered sources and the exact position of simulated sources.
 Figure~\ref{pos} shows the histogrammes of distances
 between extracted and simulated sources with fluxes equal to
 100, 200 mJy and 500 mJy for the N2 field.
 The peak of the distribution is around 8 arcsec for 100 mJy
 sources and below 5 arcsec for sources brighter than 200 mJy.

 The absolute pointing error of \ISO\ represents only a
 small additional uncertainty as it was better than a few arcsec
 all along the mission (Kessler 2000).
\subsection{Flux uncertainties}
\label{Flux_uncertainties}
\begin{figure}
\epsfig{file=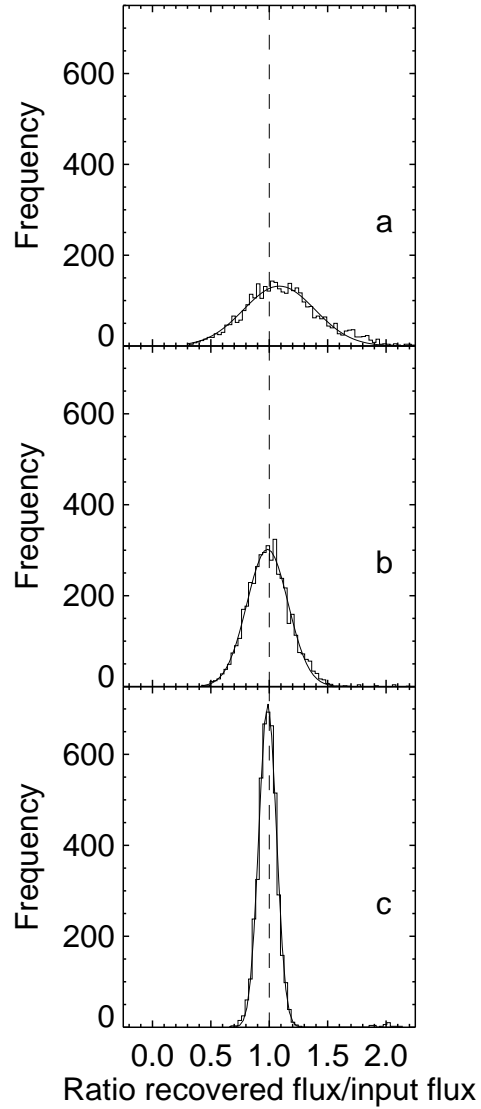,height=16cm, bb=80 80 265 540}
\caption{Histogrammes of the recovered to input flux for simulated sources
  with a flux equal to 100, 200 and 500 mJy in plots 'a','b' and 'c'
  respectively in N2. The mean unity is indicated as a dashed line. Solid
  lines are Gaussian fits to the distribution.}
   \label{gauss}
\end{figure}

 Histogrammes of the recovered to input flux of simulated sources are shown on Fig.~\ref{gauss} 
 for N2 at 100, 200 and 500 mJy.
At each flux level, the distribution 
was fitted with a Gaussian whose $\sigma$ gives an estimate of the
 photometric accuracy. Fig.~\ref{completeness_sigma}{\it b} gives the variation of $\sigma$ as
function of flux level derived from simulated sources detected with SExtractor
with a signal-to-noise $\geq 3$. 
The uncertainty on the recovered flux is typically 30 per cent at 100 mJy and
decreases to less than 10 per cent above 400 mJy.
\subsection{Completeness}
\begin{figure}
\epsfig{file=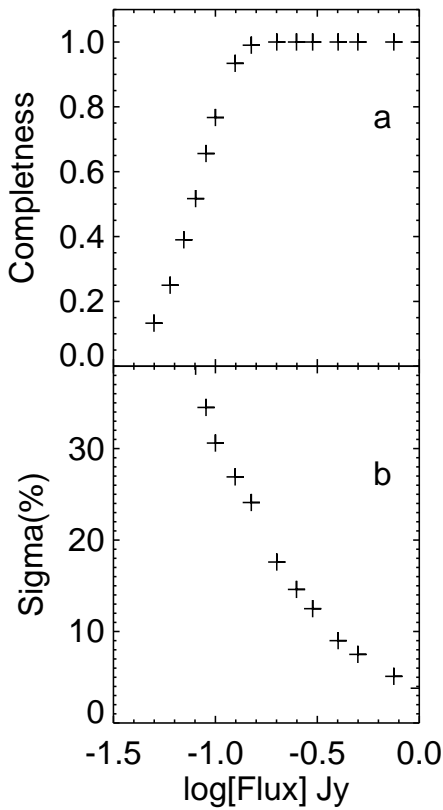,angle=0,height=11cm  , bb=100 228 260 530 }
\caption{{\it a:} Completeness of the survey as function of flux (in
  logarithmic scale) derived from
simulation of artificial sources in N2. The Completeness was computed for
sources with signal-to-noise ratio better than 3.{\it b:} Flux uncertainty 
derived from Gaussian fitting of the distribution of recovered to input
flux as function of input flux.}
\label{completeness_sigma}
\end{figure}

 The completeness of the survey is computed as the ratio of the number of 
 recovered sources with signal-to-noise ratio above 3 to the total number 
 of simulated sources and is shown on Fig.~\ref{completeness_sigma}{\it a} for N2.
 The completeness is almost 100 per cent down to 150 mJy and decreases
 to 77 per cent at 100 mJy. 

\subsection{Eddington bias}
Noise on the images is responsible for an excess in the number counts
as it will create an overestimate of fluxes.
This effect, know as Eddington bias (Eddington 1913), is similar to
Malmquist bias, which refers to fluctuation in intrinsic rather than measured
quantities (see e.g. Teerikorpi 1998). 

The proper determination of the bias plays an important role in the
estimation of source flux, the computation of number counts 
and therefore the determination of the strength of 
the evolution seen in the counts as the correction dramatically increases 
 towards the faint end of the sample.

One can estimate the Eddington bias analytically assuming a certain power-law
 and adding an appropriate flux dispersion like in e.g. Murdoch et al. 
(1973) for an underlying Euclidean slope, Oliver et al. (1995) and 
 Dole et al. (2001) who all assumed Gaussian noise. One can also use a 
 Monte-Carlo approach like Bertin, Dennefeld \& Moshir (1997).      
  
\begin{figure}
\epsfig{file=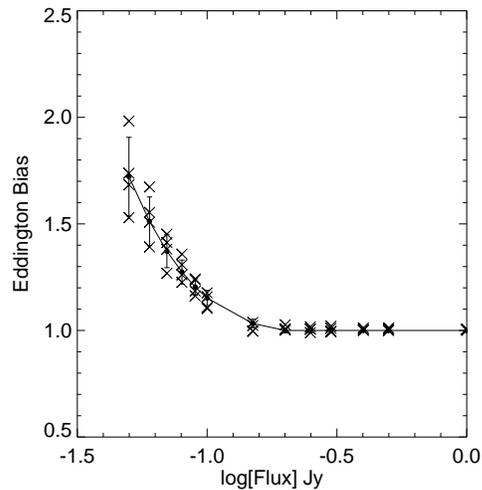,angle=0,height=7cm}
\caption{The Eddington bias as function of flux (in logarithmic scale) for
  the ELAIS survey computed as the centre of the gaussian fit to the 
 distribution of measured to input
 source flux (see Sect.~\ref{Flux_uncertainties}). The mean correction
 (diamonds) is the average of the values for the four ELAIS fields (plus signs).
 The solid line is the result of a polynomial
 fit of degree 4. Error bars are the standard deviations at each flux level.}  
\label{eddington}
\end{figure}

A more realistic estimate of the bias can be obtained from simulations
performed on the maps themselves to estimate the correction.
A mean correction of the Eddington bias was computed for the four fields and is presented in 
Fig.~\ref{eddington} as a polynomial fit to the centres of the Gaussian fits
to the distributions of measured to input flux (Sect.\ref{Flux_uncertainties} and Fig.~\ref{gauss}).
The bias is less than 34 and 13\% above 70 and 100 mJy, respectively.
The correction for the Eddington bias was directly performed on the
source flux (while the usual way is to correct the number
counts assuming a certain power law (see references above)).
This provides corrected source catalogues and does not need any
assumption on the distribution of source flux to apply the correction to the
number counts.  

\section{ELAIS 90 micron final catalogue}
\label{sec_cata}

To check the reliability of the sources detected with SExtractor, the 
classification presented in Paper III was used to check the 
reliability of the detected sources. Five persons eyeballed all the detections above  
1.5\,$\sigma$ of the sky background detected along the data streams of individual
pixels.
Only sources with detections classified at least twice as probable sources 
within a circle of 150 arcsec radius, a signal-to-noise $\geq 3$ and a flux 
$\ge$ 70 mJy (i.e. above the 3--$\sigma$ noise level computed in Sect.~\ref{confusion_noise})
were retained for the final source list.  
The selection based on the eyeball classification ensures that there are no
or few fake sources in our sample. 

The final ELAIS 90\,$\mu$m source list contains 229 sources while the 163 
most reliable sources detected in the preliminary analysis were presented 
in Paper I with flux uncertainties estimated to be 40\%.  
The comparison of source flux from the final and preliminary analysis is 
presented in Rowan-Robinson et al. (2004) for {\em ISOPHOT\/} and {\em
  ISOCAM\/} and shows a good agreement.
Table~\ref{catalogue} gives right ascension, declination, flux and flux
uncertainty (which contains the uncertainty given by SExtractor and the error coming
from the Eddington bias correction) for each source. The full version is available in electronic
format at http://www.blackwell-synergy.com.

\begin{table*}
\caption{ELAIS 90\,$\mu$m source list. Columns give the name of the source
  using the ELAISP90\_JHHMMSS+DDMMSS format according to the acronym in the
  IAU Registry, the right ascension and declination, the
flux and flux uncertainty in mJy. The full version of this table is available in electronic format at http://www.blackwell-synergy.com.}
\label{catalogue}
\begin{tabular}{lcccccccc}
\hline
Name        &          \multicolumn{3}{c}{RA (2000)}  &   \multicolumn{3}{c}{DEC (2000)}  &  S(mJy) &  e\_S(mJy)     \\
            &              h  &  m   &   s  &   deg  & '  & ''  &  mJy  &  mJy \\
\hline
ELAISP90\_J002905-432356 & 00 & 29 & 05.4 & -43 & 23 & 56.5 &  111 &   27 \\
ELAISP90\_J002915-430303 & 00 & 29 & 15.2 & -43 & 03 &  3.5 &  166 &   29 \\
ELAISP90\_J002934-431137 & 00 & 29 & 34.1 & -43 & 11 & 37.0 &  146 &   29 \\
ELAISP90\_J003000-442243 & 00 & 30 & 00.0 & -44 & 22 & 43.3 &  234 &   29 \\
ELAISP90\_J003019-424153 & 00 & 30 & 19.7 & -42 & 41 & 53.9 &   90 &   26 \\
ELAISP90\_J003023-423703 & 00 & 30 & 24.0 & -42 & 37 &  3.6 &  849 &   29 \\
ELAISP90\_J003024-433108 & 00 & 30 & 24.9 & -43 & 31 &  8.3 &  113 &   28 \\
ELAISP90\_J003032-424600 & 00 & 30 & 32.9 & -42 & 46 &  0.5 &   96 &   26 \\
ELAISP90\_J003057-441621 & 00 & 30 & 57.8 & -44 & 16 & 21.5 &  241 &   28 \\
ELAISP90\_J003059-440413 & 00 & 30 & 59.9 & -44 & 04 & 13.2 &  100 &   27 \\
ELAISP90\_J003100-435830 & 00 & 31 & 00.7 & -43 & 58 & 30.4 &  158 &   27 \\
ELAISP90\_J003105-425642 & 00 & 31 & 05.6 & -42 & 56 & 42.3 &  139 &   25 \\
ELAISP90\_J003114-431100 & 00 & 31 & 14.2 & -43 & 11 &  0.3 &  147 &   29 \\
ELAISP90\_J003124-433313 & 00 & 31 & 24.5 & -43 & 33 & 13.9 &  154 &   29 \\
ELAISP90\_J003133-424436 & 00 & 31 & 33.9 & -42 & 44 & 36.6 &  366 &   30 \\
ELAISP90\_J003135-433302 & 00 & 31 & 35.0 & -43 & 33 &  2.3 &  167 &   29 \\
ELAISP90\_J003152-440929 & 00 & 31 & 52.6 & -44 & 09 & 29.1 &  135 &   28 \\
ELAISP90\_J003218-432521 & 00 & 32 & 18.0 & -43 & 25 & 21.9 &  156 &   29 \\
ELAISP90\_J003244-423321 & 00 & 32 & 44.4 & -42 & 33 & 21.8 &  194 &   30 \\
ELAISP90\_J003249-432953 & 00 & 32 & 49.6 & -43 & 29 & 53.9 &  134 &   21 \\
ELAISP90\_J003253-424607 & 00 & 32 & 53.9 & -42 & 46 &  7.9 &  277 &   30 \\
ELAISP90\_J003300-425210 & 00 & 33 & 00.7 & -42 & 52 & 11.0 &  204 &   25 \\
ELAISP90\_J003312-423425 & 00 & 33 & 13.0 & -42 & 34 & 25.2 &   83 &   25 \\
ELAISP90\_J003316-432104 & 00 & 33 & 16.2 & -43 & 21 &  4.7 &  127 &   16 \\
ELAISP90\_J003318-440828 & 00 & 33 & 18.1 & -44 & 08 & 28.9 &  175 &   29 \\
ELAISP90\_J003321-432700 & 00 & 33 & 21.9 & -43 & 27 &  0.3 &  260 &   17 \\
ELAISP90\_J003349-441903 & 00 & 33 & 49.3 & -44 & 19 &  3.7 &   79 &   18 \\
ELAISP90\_J003359-441108 & 00 & 33 & 59.5 & -44 & 11 &  8.3 &  177 &   24 \\
ELAISP90\_J003415-423205 & 00 & 34 & 15.2 & -42 & 32 &  5.1 &   77 &   25 \\
\hline
\end{tabular}
\end{table*}


\section{Calibration comparisons}
\label{calib_comparison}
To check the quality of the calibration at the low surface brightness level
 of the ELAIS fields, we compare the {\em ISOPHOT\/} calibration with
 theoretical predictions for standard stars (Sect.~\ref{sec_stars}), and with 
\IRAS\ (Sect.~\ref{sec_iras}) flux estimates.
\subsection{Standard stars}
\label{sec_stars}

 In order to better determine
 the ELAIS calibration (as well as the general {\em ISOPHOT\/} calibration)
 three stars (HR6132, HR6464 and HR5981) close to the ELAIS fields
 were observed in mini-raster mode (a $3 \times 3$ raster with the
 star positioned at the centre of a different pixel in each
 pointing).
 The faintest of the stars (HR5981) was observed twice on the same \ISO\ orbit.

 To increase the sample of measurements and thus the reliability of the
 comparison, all other standard stars observed in mini-raster mode at 90
\,$\mu$m were retrieved from the {\em ISOPHOT\/} archive.
 The comparison with two model predictions was performed.
 Hammersley et al. (1998) models were constructed by fitting near-IR observations 
 performed with the Infrared Telescope Facility (IRTF).
 Cohen et al. (1999) constructed empirical stellar spectra in the near and 
 mid-infrared based on observations taken from the ground, the Kuiper 
 Airborn Observatory, and the \IRAS\ Low Resolution Spectrometer. 
 Both predictions were extrapolated to longer wavelengths as $\nu^2$. 

 Table~\ref{tab:stars} shows the list of stars and the characteristics
 of the {\em ISOPHOT\/} measurements as well as model predictions.
 The predicted stellar fluxes lie in the range between 60 mJy and 10Jy at 90\,$\mu$m.
 Uncertainties on the models estimates are typically 3 and 5\% for Hammersley
 et al. (1998) and Cohen et al. (1999), respectively.

%
 The integration time per pointing in these mini-rasters (from 40 to 72 s) is
 longer than that used for the bulk of the ELAIS survey in order to obtain an
 accurate determination of fluxes to establish the {\em ISOPHOT\/} calibration.

 The observations of calibration stars were processed in the same way as the survey rasters.
 The application of a method based on celestial standards
 ultimately depends on the accuracy with which the background can be
 estimated and on the accuracy of the fluxes of the sources used as
 calibrators. However, in the case of the small rasters maps performed on standard stars,
 SExtractor fails to compute a reliable sky (and therefore star flux)
 estimate.

 The best way to extract both star and sky background estimates for point
 sources observed in mini-raster mode 
 was found to be to use the individual pixel values at each raster position weighted by
 the point-spread function fractions derived for the {\em ISOPHOT\/} C100
 (Laureijs 1999, Mo\'or, in preparation).
 The fraction of the point-spread function falling on a C100 pixel situated
 at a distance $d$ from the point source centre ($f_{\rm psf}(d)$) has been determined
 for each filter at a number of typical distances. The method assumes
 the source is point-like (the $f_{\rm psf}$ factors have to be modified if the
 source is extended) and centered on the detector pixels which is the case
 for standard stars. 
{\em ISOPHOT\/} values were color-corrected
 according to the spectral type of these stars.

 Results of the comparison are given in Table~\ref{tab:stars} as the ratio
 between measured (based on the FCS) and theoretical fluxes.
 When two model predictions were available, we used their weighted mean to
 compute the ratio.
The measurement of a star (HR7451) with a very low predicted flux (7.5 mJy) was 
excluded from the comparison.
The two measurements of the brightest star (HR5340) are in very good agreement. 

{\em ISOPHOT\/} fluxes are on average higher than the predicted ones and the
weighted mean ratio is 1.06$\pm$0.02. 
It is unclear whether this discrepancy is coming from the differences in the observing
setup used for standard stars and the ELAIS survey.  
The difference between the FCS calibration and the model prediction for stars
is shown on Figs. 8, 9 and 10. 

\begin{table*}
\caption{The list of standard stars used to check the FCS calibration with theoretical values.
Colums are : the target dedicated time (TDT) number of measurement, name of stars, exposure time, 
size of the mini-rasters in steps of 46 arcsec, model predictions from Cohen
et al. 1998 ($F_{\rm MC}$) and Hammersley et al. 1998 ($F_{\rm PH}$) and
{\em ISOPHOT\/} measurements ($F_{\rm Phot}$) and their respective
uncertainties ($e_{\rm MC}$ , $e_{\rm PH}$ and  $e_{\rm Phot}$)
are indicated. ``Ratio'' is the ratio of measured to predicted
fluxes. When two model predictions were available, we used their weighted
mean to compute the ratio. 
Errors on the ratio are given in $e_{\rm Ratio}$. The weighted mean ratio is 1.06$\pm$0.02.}
\label{tab:stars}
\begin{tabular}{lccccccccccc}
\hline

                  &              &              &           &
                  \multicolumn{4}{c}{Models}             &
                  \multicolumn{2}{c}{Measurements}   & &  \\   
Measurement       &     Name     & Exposure     & Size      &   $F_{\rm MC}$   &
  $e_{\rm MC}$   &    $F_{\rm PH}$    &   $e_{\rm PH}$   & $F_{\rm Phot}$       & $e_{\rm
  Phot}$  &  Ratio & $e_{\rm Ratio}$ \\
  TDT number      &              &  sec         &   -        &     Jy      &
  Jy        &      Jy      &     Jy        &   Jy      &    Jy    &   -     &   \%    \\
\hline
 08602417         &     HR5340   &  37.00  &  5\time3   &  9.303  & 0.528 &
  9.029  & 0.300 &  9.54    &   0.34   &  1.05 & 0.05  \\
 10503417         &     HR6705   &  72.00  &  5\time3   &  2.012  & 0.115 &
  1.904 & 0.066 &  2.02    &   0.14   &   1.05 & 0.08  \\
 27502117         &     HR5340   &  72.00  &  5\time3   &  9.303  & 0.528 &
  9.029  & 0.300 &  9.52    &   0.23   &   1.05 & 0.04 \\
 29301005         &     HR7310   &  72.00  &  5\time3   &  0.258  & 0.015 &
  0.268  & 0.009 &  0.33    &   0.02   &   1.24 & 0.08 \\
 39103002         &     HR8775   &  72.00  &  5\time3   &  4.957  & 0.282 &
  5.096  & 0.184 &  5.49    &   0.15   &   1.09 & 0.04  \\
 65701318         &     HR1654   &  72.00  &  3\time5   &  0.713  & 0.042 &
  --    &     --  &  0.74    &   0.02   &   1.04 & 0.07 \\
 72701418         &     HR7980   &  72.00  &  3\time5   &  0.517  & 0.031 &
  --    &     --  &  0.48    &   0.02   &   0.93 & 0.07 \\
 77200361         &     HR5981   &  40.00   &  3\time3   &   --   & -- &
 0.063   & 0.002 &  0.07   &   0.02   &   1.11 & 0.32  \\
 77200364         &     HR5981   &  40.00   &  3\time3   &   --   & -- &
 0.063   & 0.002  & 0.07   &   0.01   &   1.11 & 0.16 \\
 78300465         &     HR6464   &  40.00   &  3\time3   &   --   & -- &
 0.120   & 0.004 & 0.13   &   0.02   &   1.08 & 0.17 \\
 78300677         &     HR6132   &  40.00   &  3\time3   &   --   & -- &
 0.288   & 0.001 & 0.32   &   0.03   &   1.11 & 0.10  \\
\hline
\end{tabular}
\end{table*}
\subsection{Comparison with \IRAS\ sources}
\label{sec_iras}
\begin{figure}
\epsfig{file=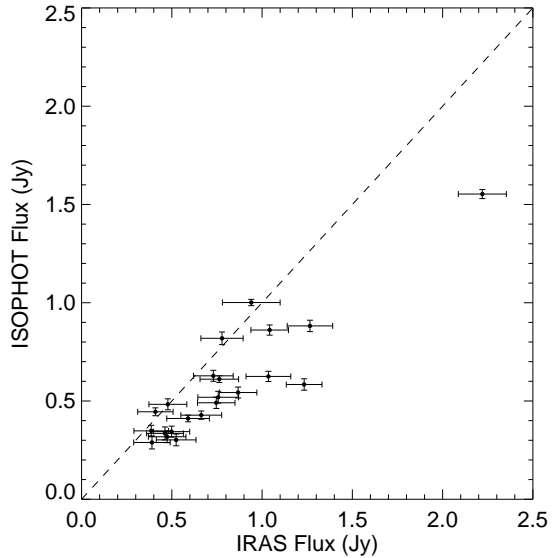 ,width=8cm}
\caption{Comparison of {\em ISOPHOT\/} and \IRAS\/FSC fluxes
at 90 micron for 21 common sources with intermediate \IRAS\ quality flags
(i.e. 2).
The 90\,$\mu$m fluxes of the \IRAS\ sources are estimated by linearly
interpolating in linear space between the colour-corrected 60 and 100\,$\mu$m fluxes.}
\label{iras_phot}
\end{figure}

While the ELAIS fields were chosen to avoid strong 12\,$\mu$m infrared sources,
there are a number of \IRAS\ 100\,$\mu$m sources detected in the survey.
All common sources have low (the flux is an upper limit) or intermediate \IRAS\
quality flags (Moshir, Kopman, \& Conrow 1992).
Figure~\ref{iras_phot} shows the comparison with the Faint Sourve Catalogue (FSC) which
is known to be more accurate than the Point Source Catalogue at faint level.
Only sources with quality flags equal to 2 (intermediate accuracy) were selected 
for the comparison and this represents 21 \IRAS\ sources.

Colour correction factors were computed from the \IRAS\ 4-band Spectral Energy 
Distribution and \IRAS\ and {\em ISOPHOT\/} filter profiles.

 The mean ratio of {\em ISOPHOT\/} to \IRAS\ flux is 0.76
 with a standard deviation of 0.17 and shows a discrepancy with the model
 prediction comparison (Sect.~\ref{sec_stars}).    

However it should be noted that there is a tendency for the \IRAS\ FSC to
overestimate fluxes near the FSC threshold at 60 and 100\,$\mu$m (Moshir et al
1992).  Since all the \IRAS\ sources in Fig.~\ref{iras_phot} have $S(100) < 3$ Jy (and most have
$S(100) < 1 Jy$), this would be sufficient to explain the discrepancy noted
above. 

\section{Correlation with FIRBACK}
\label{sec_90_170}
 We looked for FIRBACK identifications (Dole et al. 2001) of our 90\,$\mu$m 
 source sample within a circle of 188 arcsec radius (i.e. sqrt(2) $\times$
 (89.4+43.5) where 89.4 and 43.5 are the pixel size of the C100 and C200
 detectors, respectively)
, also including the complementary FIRBACK
 source catalog which provides sources with fluxes down to $135$ mJy. 
 If several sources were selected, the closest identification was used.
%
 In the common area of the two surveys, 
 53 out of 102 and 21 out of 55 FIRBACK sources
 were identified at 90\,$\mu$m in the N1 and N2 fields, respectively.

 Since {\em ISOPHOT\/} fluxes refer to a spectrum with $\nu F_{\nu} =$ constant,
 colour temperature $T_{C}$  were computed correcting the 90
 and 170\,$\mu$m fluxes in the two band-passes for a modified blackbody
 function with an emissivity index $\beta = 2$.

\begin{figure}
\epsfig{file=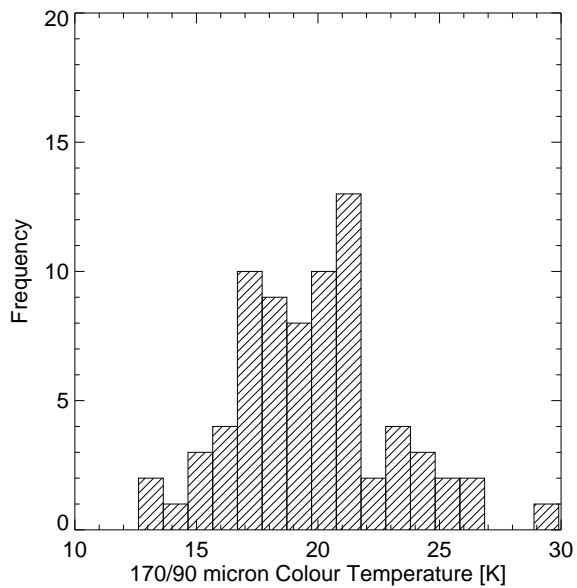,angle=0,width=8cm}
\caption{The 170\,$\mu m$/90\,$\mu$m colour temperature distribution for 53 and 21 sources detected at both wavelengths in N1 and N2 fields respectively. An emissivity index $\beta =2$ was adopted.}
\label{colour_temp}
\end{figure}

 The resulting distribution of colour temperatures
  (Fig.~\ref{colour_temp}) is centered 
 around $T_{C} = 19K$ with most of sources lying in the range 15-25K.
These values are in favour of the presence of a cold component 
in low redshift galaxies (Rowan-Robinson et al. (2004) derived a median 
redshift of 0.15 for the 90\,$\mu$m sample) which was first detected in the \ISO\ Serendipity 
Survey (Stickel et al. 1998, 2000, 2001).
Dunne and Eales (2001) also recently measured a cold component 
(20-21K) in their sample of 17 galaxies combining \IRAS\ and SCUBA 
observations at 450\,$\mu$m. 
This is consistent with the analysis of {\it COBE\/}/\DIRBE\ data for 
the MW Galaxy (Sodroski et al. 1994).
A more detailed analysis with spectral energy distribution fitting 
of the ELAIS sources from the optical to the FIR domain is presented in 
Rowan-Robinson et al. (2004).

\section{Source confusion estimates}
\label{confusion_noise} 
\begin{figure}
\epsfig{file=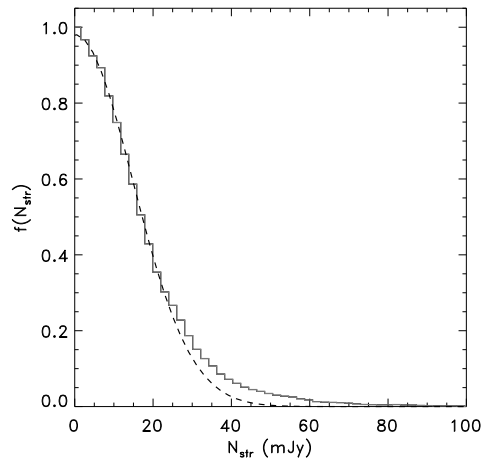,angle=0,width=7cm}
\caption[]{Distribution of the structure noise
in the ELAIS N1 field.
The solid line (histogram) represents the measured $N_{\rmn{str}}$
values over the whole field. The dashed line represents the
Gaussians fitted to the low amplitude regime (see text for details). }
\label{fig:nstr}
\end{figure}

Estimates of the confusion noise relies on the direct
measurement of structure noise, $N_{\rmn{str}}$. The structure noise
is calculated via the so-called structure function S,
which measures the average brightness fluctuations for a
specific measurement configuration see e.g. Herbstmeier et al. (1998) :
\begin{equation}
S_{\rmn{k}}(\theta) = \Big \langle \Big |
B({\underline{x}})  - {1\over{k}} \sum_{i=1}^k
B({\underline{x}} + {\underline{\theta}_{\rmn{i}}})
\Big |^2 \Big \rangle _{\rmn{x}}
\label{eq:sk}
\end{equation}
where $B({\underline{x}})$ is the measured sky brightness at the position '\underline{$x$}',
$\theta$ is the separation between the target and reference positions,
'k' is the number of reference positions,
$\underline{\theta}_{\rmn{i}}$-s are the
vectors to the reference positions relative to the target.
The average is taken over the whole map.
The actual values of $\underline{\theta}_{\rmn{i}}$-s are determined by
the geometry of the measurement configuration.
The structure noise is calculated from the structure function:
\begin{equation}
N_{\rmn{str}} = \sqrt{S_{\rmn{k}}}\times\Omega
\end{equation}
where $\Omega$ is the effective solid angle of the aperture.
The structure noise contains the contribution of the sky
brightness fluctuations (confusion noise, $N_{\rmn{conf}}$) and that of
the average instrument noise $N_{\rmn{inst}}$. As was shown by Kiss et al. (2001)
the relation between these quantities can be well approximated by
the following formula for {\em ISOPHOT\/} measurements:
\begin{equation}
N_{\rmn{str}}^2 = N_{\rmn{conf}}^2 + 2N_{\rmn{inst}}^2
\end{equation}
We derived the distribution of structure noise for 
individual pixel pairs (without averaging in space in Eq.~\ref{eq:sk})
for the ELAIS fields N1, N2, N3 and S1. The results of the ELAIS N1 field 
are presented on Fig.~\ref{fig:nstr} (the N2, N3 and S1 fields have a similar
distribution).

As seen in this figure the distribution of $N_{\rmn{str}}$ is Gaussian-like,
with an extended tail toward high structure noise values.
In brighter cirrus regions the distribution of cirrus fluctuations
can be well separated from the fluctuation of the CFIRB and the contribution
of the instrument noise. In those fields the fluctuation distribution can be 
well described by a Gaussian one (Kiss et al. 2003). In the following we assume that this distribution
can be applied for the fluctutions in the faint ELAIS fields as well.
%
Fig.~\ref{fig:nstr} can be used to estimate the source confusion
limits of the ELAIS fields. We fitted a Gaussian to the
lower noise regime ($N_{\rmn{str}}$\,$<$\,20\,mJy) of the $N_{\rmn{str}}$
distribution of the N1, N2, N3  and S1 fields, which resulted
in a $\sigma$ of 14.8, 12.8, 13.4, and 17.1\,mJy, respectively.
With the point spread function fraction coefficient of $f_{\rmn{psf}}$\,=\,0.61
for a C100 camera pixel at 90\,$\mu$m the 3--$\sigma$ source confusion limit
is $\sim$70\,mJy.
Moreover, it is worth to mention that the cosmic far-infrared background has an
expected fluctuation power of $\sim$7\,mJy at this wavelength for the
C100 camera detector pixels (Kiss et al. 2001), which contributes to
the final width of the $N_{str}$ distribution. 
Eliminating this value from the width of the Gaussians, the remaining
contributions of the cirrus fluctuations and the instrument noise would 
be $\sim$60\,mJy at 3--$\sigma$. 

\section{Source counts}
\label{sec_counts}

\subsection{ELAIS counts}%
\begin{table}
\caption{Integral number counts in the ELAIS survey. 
  S is the flux in mJy; N is the number of sources per deg$^2$ 
  with flux larger than S; 
  The upper and lower uncertainties in the counts
  which come from Poisson error and the correction of the Eddington bias are indicated.
 N\_s is the number of sources with flux larger than S.}
\label{int_counts_table_3sig}
\begin{tabular}{lcccc}
\hline
S         & N($>S$)     & N\_s  \\
mJy       & deg$^{-2}$ &        \\ 
\hline
95     &   17.02$\pm$4.50/4.30 & 185  \\  
177    &    7.93$\pm$1.19/1.83 & 90   \\
330    &    2.70$\pm$0.74/0.74 & 30  \\
614    &    0.77$\pm$0.39/0.39 & 9  \\
\hline
\end{tabular}
\end{table}
\begin{table}
\caption{Normalized differential number counts. 
  Columns are : flux range
  for each bin in mJy; the bin centre; dN/dS $\times$ S$^{2.5}$ are the
  number counts per bin normalized to the Euclidian law in
  deg$^{-2}$Jy$^{1.5}$ 
  (the bin centre in linear scale was used for the normalisation); 
  The upper and lower uncertainties in the counts
  which come from Poisson error and the correction of the Eddington bias are indicated.
  N\_s is the number of sources per bin.}
\label{dif_counts_table_3sig}
\begin{tabular}{lccc}
\hline
flux bin   &  bin centre   & dN/dS $\times$ S$^{2.5}$  & N\_s  \\
mJy        &     mJy       & deg$^{-2}$ Jy$^{1.5}$     &       \\ 
\hline
   95--176  &   135  &  0.74$\pm$0.20/0.22  & 95 \\
  176--329  &   253  &  1.11$\pm$0.24/0.24  & 60\\
  329--613  &   471  &  1.04$\pm$0.27/0.27  & 21\\
  613--1142 &   877  &  0.85$\pm$0.32/0.32  & 7\\
\hline
\end{tabular}
\end{table}

Integral and normalized differential source counts 
are given in Tables~\ref{int_counts_table_3sig} and
\ref{dif_counts_table_3sig} for 185 sources brighter than 95mJy (above this
flux, the
uncompleteness and Eddington bias corrections are lower than 25 and 15\% respectively). 
Uncertainties in the counts represent the contribution of Poisson errors and the Eddington bias correction.
The possibility that some sources could be solar system bodies was rejected
in Sec.\ref{sec_extraction}. 
 Moreover as stated in Paper III, given that there are no bright
 12-$\mu$m sources in the ELAIS fields (to avoid saturating ISOCAM) we do not
 expect any 'photospheric' stars to be detected at 90\,$\mu$m as these would
 have a 90\,$\mu$m flux $\leq$ 10 mJy. 
Finally, the 3--$\sigma$ limit of the cirrus and instrumental noise was estimated in
Sect. \ref{confusion_noise} to be $\sim$60 mJy.
Therefore it is very likely that all the selected sources above 95 mJy are extragalactic.

\begin{figure}
\epsfig{file=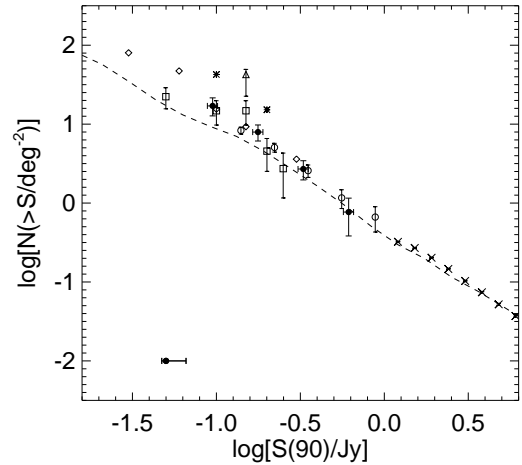,angle=0,width=8cm}
\caption{Integral source counts at 90\,$\mu$m for ELAIS (filled circles). Error
  bars are Poisson error plus the uncertainties on the Eddington bias.
Flux uncertainties are 7\% resulting from the use of the FCS for
 calibration purposes (Sect. 2.3).
  Correction factors of 1/1.06 and 1/0.76 derived from the comparison of the 
  FCS calibration with standard stars model predictions and with \IRAS\
  respectively are shown as -0.03 dex and +0.12 dex error bars at (-1.3,-2.0).  
Source counts from Juvela et al. (2000; asterisks), the preliminary analysis of ELAIS 
(Paper I; open circles), Linden-V{\o}rnle et al. (2000; squares), Matsuhara et al. 
(2000; triangle) and Rodighiero et al. (2003; diamonds) are shown for comparison.
 \IRAS\ counts (x symbols) are shown for galaxies in the PSCz catalogue. 
 The dashed line is the no-evolution model from Franceschini et al. (2001).}
\label{int_counts_with_iras}
\end{figure}

Figure~\ref{int_counts_with_iras} shows ELAIS integral counts (filled
circles) at 90\,$\mu$m 
compared with results of Juvela et al. (2000; asterisks), the preliminary analysis 
of ELAIS (Paper I; open circles), Linden-V{\o}rnle et al. (2000; squares), 
Matsuhara et al. (2000) from the Lockman Hole observations (triangle) and the
new analysis of the Lockman Hole performed by Rodighiero et al. (2003) (diamonds).
\IRAS\ points (x symbols) are also shown for galaxies in the PSCz catalogue (Saunders et al. 2000)
 with a selection of galactic latitude ($|b| > 20$) and low \IRAS\ 
 flags ($f_{\rm qual} < 3$) at 100\,$\mu$m and fluxes brighter than 2Jy as 
 it becomes incomplete at fainter level (see Paper III for details). 
The dashed line is the no-evolution model from Franceschini et al. (2001).
  Correction factors of 1/1.06 and 1/0.76 derived from the comparison of the 
  FCS calibration with standard stars model predictions
  (Sect.~\ref{sec_stars}) and with \IRAS\ (Sect.~\ref{sec_iras}) are also shown.

Our results extend \IRAS\ counts by more than one order of magnitude. They are in 
very good agreement with the preliminary analysis of ELAIS and confirm the 
departure from the Euclidian slope found in Paper I. 
Integral counts  in the range 0.095-1Jy are well fitted with a straight line of the form:
\begin{equation}
       \log_{10}(N) = (-1.68\pm 0.09)  \times \log_{10}(S\mathrm{[Jy]}) - (0.43\pm0.07)
\end{equation}
    
\subsection{Comparison with evolutionary models}
\label{model_comparison}
\begin{figure}
\epsfig{file=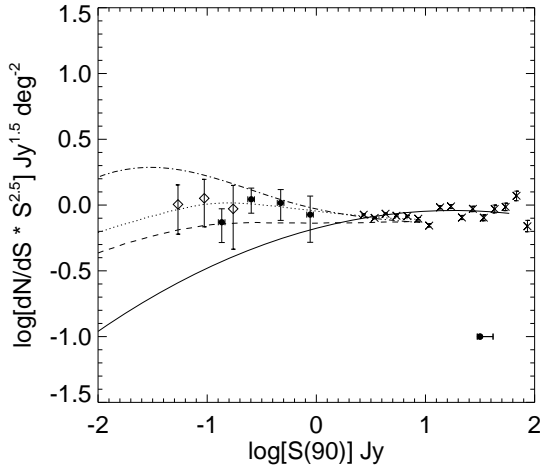,angle=0,width=8cm}
\caption{Normalized differential source counts for ELAIS (filled circles) at
  90\,$\mu$m. Error bars include Poisson error and the uncertainties on the
  Eddington bias. 
  Correction factors of 1/1.06 and 1/0.76 
 are shown as -0.03 dex and +0.12 dex error bars at (1.5,-1.0).  
Lockman Hole counts from Rodighiero et al. (2003) are also shown (diamonds). \IRAS\ counts (crosses) are shown for galaxies in the PSCz catalogue.
 The solid line is the no-evolution model from Franceschini et al. (2001). 
 The dashed and dotted line represent models A and E of Guiderdoni et al. 
(1998), respectively. The dash-dotted line are the counts predicted by 
Rowan-Robinson (2001) for his cosmological model with $\Omega_0=0.3$ and 
$\Lambda=0.7$.}
\label{diff_counts_figure1}
\end{figure}
\begin{figure}
\epsfig{file=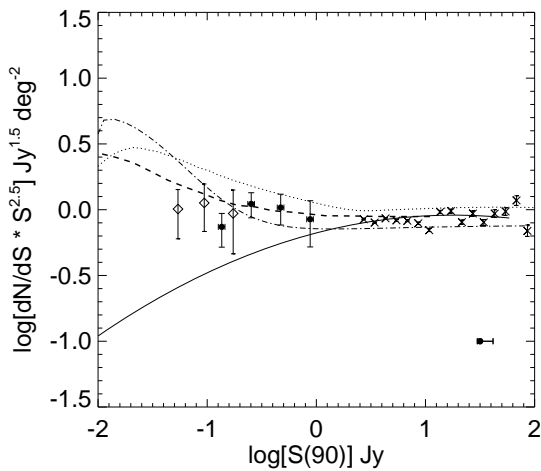,angle=0,width=8cm}
\caption{Same as \ref{diff_counts_figure1} but the dashed-dotted, dashed and 
dotted lines are the models of Pearson (2001), Franceschini
 et al. (2001) and Lagache, Dole \& Puget (2003), respectively.}  
\label{diff_counts_figure2}
\end{figure}

 Recent observations in the FIR and sub-millimeter regimes have considerably 
 improved evolutionary models in the past 5 years.
 In the following, we briefly describe the main characteristics of   
 evolutionary models by Guiderdoni et al. (1998), Rowan-Robinson (2001),  
 Pearson (2001), Franceschini et al. (2002), Lagache et al.
 (2003) and compare the predictions to differential number counts 
 measured in the ELAIS survey on Figs.~\ref{diff_counts_figure1} and \ref{diff_counts_figure2}. 
\begin{enumerate}

\item Guiderdoni et al. 1998 have designed a family  of
semi-analytic evolutionary scenarios within the context of 
hierarchical growth of structures according to the cold dark matter model,
with prescriptions for dissipative and non-dissipative collapses, star
formation and feedback. Differences between these scenarios only 
concern the efficiency of star formation on a dynamical time-scale, the IMF
and the extinction. 
In Figure~\ref{diff_counts_figure1} we compare our results with two of 
their models:

\begin{description}
\item Scenario A contains a mix of two broad types of populations, one with a
  'quiescent' star formation rate, the other proceeding in bursts 
   with a high evolution rate and fitting the SFR density at low z;
\item Scenario E includes an additional population of heavily-extinguished 
galaxies (ULIGs) 
 and is qualified as the best fit by Guiderdoni et al.
as it nicely reproduces the Cosmic Optical Background and the Cosmic Infrared Background.
\end{description}

If model A is systematically below the measured counts, the addition of a
ULIGs population shifts the predictions upwards and the model E falls in
excellent agreement with the observations as it was suggested in Paper III 
based on the brightest sources.
\item The models of Rowan-Robinson (2001) include four spectral components:
infrared cirrus, an M82-like starburst, an Arp 220-like starburst, and an
active galactic nucleus dust torus. The proportion of each spectral type are
chosen for consistency with \IRAS\ and SCUBA colour-luminosity relations and
with the fraction of AGNs as a function of luminosity in 12\,$\mu$m samples.

 The prediction of the Rowan-Robinson model for the cosmological model 
 with $\Omega_0=0.3$ and $\Lambda=0.7$ is compared with the observed counts 
 on Figure~\ref{diff_counts_figure1}. The model is in good agreement with
 the observations down to fluxes of $\sim 200$ mJy within the error bars. 
 At fainter fluxes it gives a slightly too high number of sources.

\item The model of Pearson and Rowan-Robinson (1996) consists of non-evolving
spiral and elliptical components mixed with an evolving population of
starburst galaxies, active galactic nuclei and a hyperluminous galaxy
component. The model is in agreement with the source counts at 60\,$\mu$m 
and the faint radio counts at 1.4 Ghz and provides a good estimate of the
cosmic infrared background observed with {\it COBE\/} at 500\,$\mu$m. 

More recently, Pearson (2001) used the framework of the Pearson and
Rowan-Robinson galaxy evolution model and constrains the evolution in 
the galaxy population with the observed counts and background measurement 
derived from \ISO\ and SCUBA observations. Pearson found that a strong 
evolution in both density and luminosity of the ULIG population can account
for the source counts from 15\,$\mu$m to the submillimetre region, as well as
explain the peak in cosmic infrared background at $\sim$ 140\,$\mu$m.   

The prediction for this model is also shown on Figure~\ref{diff_counts_figure2}.
The model provides a good fit to the ELAIS observations although it seems to become
too high at fluxes fainter than $\sim$ 100 mJy compared to the Lockman Hole
counts of Rodighiero et al. 2003.

\item The model of Franceschini et al. (2001) assumes
 that the extragalactic population is composed of three components with
 different evolution properties: (1) a non-evolving population of spirals; (2)
 a population of strongly evolving starburst galaxies and type-II AGNs; (3) a
 population of type-I AGNs which does not contribute significantly to the counts. 
 This model was optimized to reproduce the mid-IR counts and redshift
 distribution. In particular, the two components of the fast evolving 
 population were required to reproduce the shape of the $15 \mu$m counts.  

  The model of Franceschini is plotted on
 Figure~\ref{diff_counts_figure2} and gives a good estimate of 
 the observed counts.
 The increase in number counts seen at fluxes fainter than 
 $\sim 100$ mJy is probably the far-infrared counterpart of the upturn 
 detected in the mid-infrared (Elbaz et al. 1999, Chary and Elbaz 2001,
 Mazzei et al. 2001; Serjeant et al. 2000). The ELAIS 90\,$\mu$m data do not allow 
 us to test if this predicted feature is real or not but Rodighiero et al. (2003) 
 have shown it is compatible within the error bars with the faint 
 counts in the ``Lockman Hole''.     
\item Lagache, Dole \& Puget (2003) have developed a phenomenological model 
which fits all the existing counts and redshift distributions from the 
mid-infrared to the submillimetre range together with the intensity and 
fluctuation of the cosmic infrared background. 
Their model is based on the evolution of galaxy luminosity function with
redshift for a population of starburts and normal galaxies.  

The model of Lagache, Dole \& Puget 2003 (shown on
Figure~\ref{diff_counts_figure2} is compatible 
with ELAIS counts (although slightly higher) around 1 Jy and
with a larger discrepancy at fainter level ($ S \leq 200$ mJy) where the model 
continues to increase while the observed counts decrease. 
\end{enumerate}

\section{Summary and discussion}
\label{sec_conclusion}
  We have used a new method to reduce {\em ISOPHOT\/} measurements in the 4
 main areas of the ELAIS survey at 90\,$\mu$m. 
 With a total area of more than 12\,deg$^2$, the ELAIS survey represents the
 largest area covered in a single programme with \ISO. 
 
 The relative uncertainty in flux coming from the FCS calibration estimated 
from the sky background level differences of all rasters is 7 per cent.

 On the one hand, the comparison of measured fluxes with models for standard stars
 shows a strong correlation with a mean ratio of {\em ISOPHOT\/} to model
 values of $1.06\pm0.02$. 
On the other hand, the comparison with the \IRAS\/FSC catalogue 
 for \IRAS\ sources detected in the survey gives a mean ratio of {\em ISOPHOT\/} to \IRAS\
 values equal to 0.76$\pm$0.17.
%

 Simulations of artificial sources on the final maps spanning a wide range of
 flux were used to estimate flux and positional uncertainties, completeness and 
 the Eddington bias corrections.
 The completeness of the survey is about $80\%$ at 100 mJy.

We present a source list of 237 reliable sources with fluxes larger than 70
mJy, signal-to-noise $\geq 3$ 
 for the 4 large ELAIS fields. 
The full version of the catalogue is available at http://www.blackwell-synergy.com.

 Sources detected at 90 and 170\,$\mu$m in the FIRBACK survey (Dole et al. 2001) 
 have an average colour temperature of $T_{C} = 19K$ with all sources lying in 
 the range 13-25K in agreement with
 Stickel et al. (2001) in the \ISO\ Serendipity Survey (Stickel et al. 1998).

 The ELAIS counts extend the \IRAS\ counts by more than one order of magnitude in
 flux and show significant departure from the no-evolution model as detected 
 in other \ISO\ surveys from the mid- to the far-infrared. 
 
 There is in general a good agreement between ELAIS and other 90\,$\mu$m source
 counts and in particular with the deeper counts measured 
 in the Lockman Hole (Rodighiero et al. 2003).
 Differential number counts measured in the ELAIS regions at 90\,$\mu$m are 
 compared to recent evolutionary models. Among few models which were
 compared to our counts, the model of Franceschini et
 al. (2001) and the scenario E of Guiderdoni et al. (1998)
give the best agreement with the observations.

 However, the latter model is a factor  of $\sim 2.5$ {\it below} the counts 
 measured at 170\,$\mu$m in two of the ELAIS regions (Dole et al. 2001) 
 related to the present paper. 
 On the other hand Matsuhara et al. (2000) 
 found that the scenario E model prediction of Guiderdoni et al. is in close
 agreement with the 170\,$\mu$m number counts in the small area of the 
 ``Lockman Hole'' (see also Kawara et al. 1998) but their 90\,$\mu$m
 integral counts are significantly above the model.
 
 The nature and redshift distributions of the ELAIS galaxies can test
 the various models and their hypothesis e.g. distinguishing the different
 galaxy populations on which these models are built.
 This will also help to clarify the origin of the differences seen in the
 number counts of the various \ISO\ surveys at different wavelengths.

 The 90\,$\mu$m luminosity function is presented in Serjeant et
 al. (2004) and Rowan-Robinson et al. (2004) present results based on the ELAIS final-band
  merged catalogue combining the \ISO\ and ground-based observations in 
  the ELAIS fields.

\section{Acknowledgments}
 We thank the referee Yasunori Sato for detailed comments and suggestions which helped us improve the manuscript.
 It is a pleasure to acknowledge Peter \'Abrah\'am and Ulrich Klaas for very useful
 discussions about {\em ISOPHOT\/} calibration and data analysis. 
  We also thank Guilaine Lagache and Michael Linden-V{\o}rnle   
 for making their results available to us in electronic version. 
 We are grateful to Emmanuel Bertin for valuable advice on the use of SExtractor.

Ph. H\'eraudeau acknowledges support from the EU TMR Network ``SISCO'' (HPRN-CT-2002-00316).

C. del Burgo acknowledges support from the EU TMR Network ``POE'' (HPRN-CT-2000-001380).

 ELAIS was supported by EU TMR
 Network FMRX-CT96-0068 and PPARC grant GR/K98728.

 This paper is based on observations with \ISO\, an ESA project with
 instruments funded by ESA member states (especially the PI countries:
 France, Germany, the Netherlands and the United Kingdom) and with
 participation of ISAS and NASA. The {\em ISOPHOT\/} data were  processed
 using PIA, a joint development by the ESA Astrophysics Division and
 the {\em ISOPHOT\/} consortium led by MPI f\"{u}r Astronomie, Heidelberg.
 Contributing Institutes are DIAS, RAL, AIP, MPIK, and MPIA.

The development and operation of {\em ISOPHOT\/} were supported by MPIA and funds
from Deutsches Zentrum f\"ur Luft- und Raumfarht (DLR, formerly DARA). The
{\em ISOPHOT\/} Data Center at MPIA is supported by Deutsches Zentrum f\"ur Luft-
und Raumfarht e.V. (DLR) with funds of Bundesministerium f\"ur Bildung und
Forschung, grant No. 50 QI 9801 3.

\label{lastpage}

\end{document}